\documentclass[onecolumn,sort&compress,numbers]{els-mrw} 

\usepackage{amsmath,amssymb,amsfonts,amsthm,graphicx}
\usepackage{txfonts}
\usepackage{helvet}
\usepackage{hyperref}
\usepackage{slashed}

\newcommand{\mpi}{M_\pi}
\newcommand{\mN}{m_N}
\newcommand{\Fpi}{F_\pi}
\newcommand{\MeV}{\,\text{MeV}}

\begin{document}


\chapter{Nucleon mass: trace anomaly and $\boldsymbol{\sigma}$-terms}\label{chap1}

\author[1]{Martin Hoferichter}%
\author[2]{Jacobo Ruiz de Elvira}%

\address[1]{Albert Einstein Center for Fundamental Physics, Institute for Theoretical Physics, University of Bern, Sidlerstrasse 5, 3012 Bern, Switzerland}
\address[2]{Universidad Complutense de Madrid, Departamento de F\'isica Te\'orica and IPARCOS,
Facultad de Ciencias F\'isicas, Plaza de las Ciencias 1, 28040 Madrid, Spain}

\maketitle

\begin{abstract}[Abstract]
We give a pedagogical introduction to the origin of the mass of the nucleon. We first review the trace anomaly of the energy-momentum tensor, which generates most of the nucleon mass via the gluon fields and thus contributes even in the case of vanishing quark masses. We then discuss the contributions to the nucleon mass that do originate from the Higgs mechanism via the quark masses, reviewing the current status of nucleon $\sigma$-terms that encode the corresponding matrix elements.  
\end{abstract}

\begin{keywords}
 	nucleon mass\sep trace anomaly\sep $\sigma$-terms
\end{keywords}

\begin{figure}[h]
	\centering
	\includegraphics[width=7.3cm]{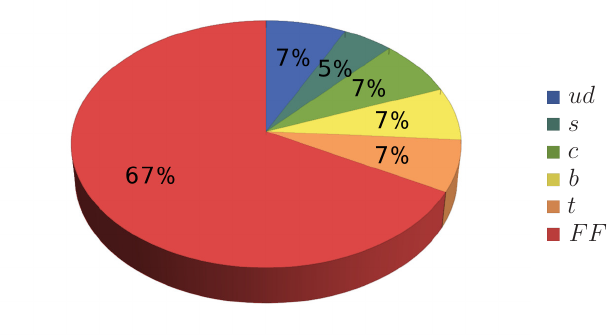}
	\caption{Pie chart for the decomposition of the nucleon mass into contributions arising from the light quark masses $m_q$ with $q=u,d$ and $q=s$, the heavy quarks $Q=c,b,t$, and the gluon field strength $F_{\mu\nu}^a F^{\mu\nu}_a$. The numerical values, taken from Eq.~\eqref{mN_numerics}, in some cases still carry substantial uncertainties, as discussed in this chapter, while the general hierarchy among the different contributions is robust.}
	\label{fig:mN_decomposition}
\end{figure}

\section*{Objectives}
\begin{itemize}
	\item Decomposition of the nucleon mass: how does the mass of the nucleon arise, and how large are the respective contributions?
	\item Trace anomaly: how does the energy stored in the gluon fields contribute to the nucleon mass?  
	\item $\sigma$-terms: how large are the quark-mass contributions to the nucleon mass and how can their values be determined?
\end{itemize}

\section{Introduction}
\label{intro}

The masses of hadronic states are far more complicated than the masses of leptons, since several mechanisms can play a role for composite states. First, the masses of the quarks contribute, generating part of the overall mass of the hadron, but also the gluon field strength matters for mass generation. For instance, in the case of the pion, its mass squared, $\mpi^2$, is directly proportional to the quark masses according to the Gell-Mann--Oakes--Renner relation~\cite{Gell-Mann:1968hlm}
\begin{equation}
\label{GOR}
 \mpi^2=2B\hat m,\qquad \hat m=\frac{m_u+m_d}{2},\qquad
 B=-\frac{1}{2F_\pi^2}\langle 0|\bar u u+\bar d d|0\rangle,
\end{equation}
where $\hat m$ is the average of the up- and down-quark masses and the coefficient $B$ can be expressed in terms of the pion decay constant $\Fpi$ and the quark condensate $\langle 0|\bar u u+\bar d d|0\rangle$. On the other hand, there exist states in the QCD spectrum whose mass originates (almost) exclusively from the gluon fields, so-called glueballs~\cite{Morningstar:1999rf,Klempt:2007cp}. The mass of the nucleon, $\mN$, is an interesting case in between, in that a large part of its mass is generated by the gluons, while quark-mass contributions are still important. Moreover, the quantification of the size of the different contributions to the mass is intimately related to nucleon matrix elements of the scalar current, whose knowledge is required in the search for physics beyond the Standard Model in a variety of processes, including dark matter in direct-detection experiments~\cite{Bottino:1999ei,Bottino:2001dj,Ellis:2008hf,Hoferichter:2016nvd,Hoferichter:2017olk,Hoferichter:2018acd},
lepton flavor violation in $\mu\to e$ conversion in
nuclei~\cite{Cirigliano:2009bz,Crivellin:2014cta,Cirigliano:2022ekw,Davidson:2022nnl,Hoferichter:2022mna,Heinz:2024cwg}, neutrino scattering~\cite{Altmannshofer:2018xyo,Hoferichter:2020osn}, and
electric dipole
moments~\cite{Engel:2013lsa,deVries:2015gea,deVries:2016jox,Yamanaka:2017mef}.

Formally, the separation between quark-mass and gluon contributions to the mass of a hadron proceeds via the trace of the energy-momentum tensor $\theta_{\mu\nu}$. That is, as we will discuss in Sec.~\ref{sec:trace_anomaly}, one has  $\langle H|\theta^\mu_\mu|H\rangle = 2M_H^2$ for a hadron $H$, so that with a suitable normalization of states one can express its mass $M_H$ in terms of matrix elements of $\theta^\mu_\mu$. For instance, for the pion one has the decomposition
\begin{equation}
 \langle \pi(p')|\theta_{\mu\nu}|\pi(p)\rangle =\frac{1}{2}\big(g_{\mu\nu}q^2-q_\mu q_\nu\big)\theta_1(q^2)+\frac{1}{2}P_\mu P_\nu \theta_2(q^2),\qquad P_\mu=p_\mu'+p_\mu,\qquad q_\mu = p_\mu'-p_\mu,
\end{equation}
so that the leading-order prediction in chiral perturbation theory (ChPT), $\theta_1(q^2)=\theta_2(q^2)=1$~\cite{Donoghue:1991qv}, indeed gives
\begin{equation}
 \langle \pi|\theta^\mu_\mu|\pi\rangle\equiv
  \langle \pi(p)|\theta^\mu_\mu|\pi(p)\rangle = \bigg[\frac{3}{2}q^2\theta_1(q^2)+\frac{1}{2}\big(4\mpi^2-q^2\big)\theta_2(q^2)\bigg]_{q^2=0}=2\mpi^2.
\end{equation}
While at the classical level, scale invariance ensures that $\theta^\mu_\mu=0$ in the absence of quark masses, quantum corrections from the gluon fields make the trace non-vanishing, and these can induce substantial contributions to the mass as well. As for the quark-mass effects, their direct contribution can be inferred via the Feynman--Hellmann theorem~\cite{Feynman:1939zza,Hellmann:1937,Gasser:1979hf}, from which it follows that the matrix element of the scalar current $m_q \bar q q$ for a quark $q$ can be expressed as
\begin{equation}\label{eq:FH_def}
 \frac{1}{2M_H}\langle H|m_q\bar q q|H\rangle =m_q\frac{\partial M_H}{\partial m_q}.
\end{equation}
In particular, for the pion, one has by virtue of Eq.~\eqref{GOR}
\begin{equation}
\label{M_pi_sigma}
\sigma_\pi\equiv\frac{1}{2\mpi}\langle\pi|m_u\bar u u+m_d\bar d d|\pi\rangle = m_u\frac{\partial \mpi}{\partial m_u}+m_d\frac{\partial \mpi}{\partial m_d}=\hat m\frac{\partial \mpi}{\partial \hat m}=\frac{\mpi}{2},
\end{equation}
i.e., for the pion, the direct so-called ``$\sigma$-term'' contribution to the mass, as quantified by the matrix element of the scalar current, gives exactly half the mass. Of course, $\mpi$ vanishes in the chiral limit $m_q\to 0$, but only half its value comes from $\theta^\mu_\mu\big|_{m_q}\equiv \sum_q m_q\bar q q $, while the remainder comes from the matrix element of the rest of $\theta^\mu_\mu$, as discussed in Sec.~\ref{sec:trace_anomaly}.
Similarly, we will dissect the mass of the nucleon in Sec.~\ref{sec:mass_nucleon}, to quantify its $\sigma$-term contributions and the finite remainder in the chiral limit that purely corresponds to gluonic effects. A summary of the current understanding of the resulting decomposition is provided in Sec.~\ref{sec:summary}.

\section{Trace anomaly of the energy-momentum tensor}\label{sec:trace_anomaly}

\subsection{General considerations}

In classical field theory, scale invariance implies that the equations of motion---and thus all physical predictions---remain unchanged under the uniform rescaling
$x^\mu\to\lambda x^\mu$ of the coordinate system as long as the theory does not contain any fixed mass or length scale, e.g., QCD with $m_q=0$. One important consequence is that the trace of the energy-momentum tensor vanishes, $\theta^\mu_\mu=0$.
However, in quantum field theory, the regularization and renormalization of ultraviolet (UV) divergences necessarily introduce a renormalization scale $\mu$. Even if the classical Lagrangian ${\cal L}$ contains no dimensionful parameters, the running of the coupling constants with~$\mu$ breaks scale invariance at the quantum level, inducing a finite trace $\theta^\mu_\mu\neq 0$.

Classically, $\theta^\mu_\mu=0$ arises as follows: the energy-momentum tensor is a conserved Noether current associated with invariance under spacetime translations $x^\mu\to x^\mu+a^\mu$, thus $\partial_\mu \theta^{\mu\nu}=0$. Second, the Noether current associated with the rescaling $x^\mu\to\lambda x^\mu$ is the so-called dilatation current $d^\mu=x_\nu\theta^{\mu\nu}$, and the combination of both symmetries then implies $0=\partial_\mu d^\mu=\theta^\mu_\mu+x_\nu\partial_\mu \theta^{\mu\nu}=\theta^\mu_\mu$.

Even in the absence of fixed length scales, scale invariance is broken by quantum effects. In practice, this is caused by the necessity to regulate UV divergences in the calculation of quantum corrections, which leads one to define renormalized couplings $g_i(\mu)$ at the arbitrary renormalization scale $\mu$. Even after removing the regulator, physical amplitudes may retain a residual dependence on $\mu$ due to higher orders in perturbation theory, signaling quantum breaking of classical scale invariance. This scale dependence is parameterized by beta functions and anomalous dimensions,
\begin{equation}
    \beta_{g_i}=\mu\frac{\text{d} g_i}{\text{d}\mu},\quad\gamma_{ij}=(Z^{-1})_{ik}\,\mu\frac{\text{d}Z_{kj}}{\text{d}\mu},
\end{equation}
where $\beta_{g_i}$ governs the running of the coupling $g_i$. $\gamma_{ij}$ is the anomalous dimension matrix for operators $O_i$ with renormalization constants $Z_{ij}$, i.e., since the bare operators $O_i^{(0)}=Z_{ij} O_j$ do not depend on $\mu$, the derivative of the renormalized operators becomes
$\mu\frac{d O_i}{d\mu}=-\gamma_{ij} O_j$, and likewise for the coefficients $\mu\frac{d c_i}{d\mu}=\gamma_{ji} c_j$, since $c_i=Z_{ji}c_j^{(0)}$.
Accordingly, in the quantum theory, the naive conservation of the dilatation current acquires an extra term due to the $\mu$-dependence of correlators, 
which leads to the master formula
\begin{equation}\label{eq:vacuum-trace}
\theta^\mu_{\mu}= \sum_i\beta_{g_i}\,\frac{\partial\mathcal{L}}{\partial g_i} - \sum_{ij}\gamma_{ij}\,O_j,
\end{equation}
where the sums run over all couplings and composite operators present in the theory. This relation makes explicit how the beta functions $\beta_{g_i}$ and anomalous dimensions $\gamma_{ij}$ generate the trace anomaly.
To relate matrix elements of $\theta^\mu_\mu$ to hadron masses, we observe that by Lorentz invariance $\langle H(p)|\theta^{\mu\nu}|H(p)\rangle\propto p^\mu p^\nu$, so that  $\langle H(p)|\theta^{\mu}_\mu|H(p)\rangle\propto M_H^2$, and the proportionality constant is determined from the component $\theta^{0\mu}$, whose matrix element gives a particle's momentum. Comparing the $\mu=0$ component then shows that for relativistically normalized states, $\langle \mathbf{p}'|\mathbf{p}\rangle=2E_p\delta^{(3)}(\mathbf{p}'-\mathbf{p})$, one indeed has $\langle H|\theta^\mu_\mu|H\rangle=2M_H^2$.

Finally, when the Lagrangian contains an explicit mass term, e.g., for a fermion field $\psi$ of renormalized mass mass $m=Z_m m^{(0)}$, the general formula~\eqref{eq:vacuum-trace} includes an explicit breaking of scale invariance
\begin{equation}
\frac{\partial{\cal L}}{\partial m}=-\bar\psi\psi,\quad \gamma_m=\mu\frac{\text{d}\log m}{\text{d}\mu}=\frac{\mu}{Z_m}\frac{dZ_m}{d\mu},
\end{equation}
so that the trace picks up the term
\begin{equation}
    \theta^\mu_\mu\supset m\left(1-\gamma_m\right)\bar\psi\psi.
\end{equation}

\subsection{Abelian case}

Having established how mass arises from the energy-momentum tensor, we now consider QED as a specific example
\begin{equation}
    {\cal L}_{\text{QED}}=\bar\psi (i\slashed D-m) \psi-\frac{1}{4} F_{\mu\nu} F^{\mu\nu},\qquad D_\mu=\partial_\mu+ieA_\mu,\qquad F_{\mu\nu}=\partial_\mu A_\nu-\partial_\nu A_\mu,
\end{equation}
where $\psi$ is the Dirac spinor field, $m$ its mass, $A_\mu$ the vector gauge field associated to the photon, and $F_{\mu\nu}$ the field strength tensor. The massless Lagrangian is indeed scale invariant, so that classically the only contribution arises from the explicit mass term, $\theta^\mu_\mu\big|_\text{classical}=m\bar\psi\psi$.  At the quantum level, however, one finds
\begin{equation}
\theta^\mu_\mu\big|_\text{QED}=\beta_{e}\,\frac{\partial\mathcal{L}_\text{QED}}{\partial e} + \beta_{m}\,\frac{\partial\mathcal{L}_\text{QED}}{\partial m},\quad\text{with}\quad \beta_e=\mu\frac{\partial e}{\partial\mu},\quad \beta_m=\mu\frac{\partial m}{\partial\mu}=m\,\gamma_m,
\end{equation}
so that~\cite{Adler:1976zt}
\begin{equation}
    \theta^\mu_{\mu}\big|_\text{QED}=\frac{\beta_e}{2e}F_{\mu\nu}F^{\mu\nu}+m\left(1-\gamma_m\right)\bar\psi\psi.
\end{equation}
The first term is the trace anomaly proportional to the running of the gauge coupling, it can be understood heuristically by absorbing the gauge coupling into rescaled fields $\bar A_\mu=e A_\mu$ and taking the derivative with respect to the remaining explicit factors of $e$. 
The second term combines both the classical breaking of scale invariance due to the fermion mass with the anomalous dimension $\gamma_m$ of the corresponding operator. At leading order,
with $\beta_e=\frac{e^3}{12\pi^3}+{\cal O}(e^5)$, $\gamma_m={\cal O}(e^2)$, this equation simplifies to
\begin{equation}
    \theta^\mu_\mu\big|_\text{QED}=\frac{e^2}{24\pi^3}F_{\mu\nu}F^{\mu\nu}+m\,\bar\psi\psi+{\cal O}\left(e^4,e^2\,m\right).
\end{equation}

\subsection{Non-Abelian case}

For QCD, the previous discussion needs to be extended to the non-Abelian case, including the self-interactions among the gluon fields, giving rise to two of the most striking phenomena in modern particle physics: \emph{asymptotic freedom} at high energies and \emph{confinement} at low energies.  As a result, the trace anomaly in non‑Abelian theories acquires a new structure, with pure‑gluon contributions running with the gauge coupling and combining with fermion mass terms to drive the dynamics of hadronic states.
In this section, we therefore generalize our previous Abelian discussion to Yang--Mills theory with gauge group $G$, gauge fields $A_\mu^a$, and $N_f$ fermion fields $\psi_f$ in representation $R$
\begin{equation}
  \mathcal{L}_{\rm YM}
  =
    \sum_{f=1}^{N_f}\bar\psi_f\,(i\slashed{D}-m_f)\,\psi_f-\frac{1}{4}\,F^a_{\mu\nu}F^{\mu\nu}_a,
  \qquad
  F^a_{\mu\nu}
  = \partial_\mu A^a_\nu - \partial_\nu A^a_\mu + g f^{abc}\,A^b_\mu A^c_\nu,
\end{equation}
where
$D_\mu = \partial_\mu + igT^aA^a_\mu$, with generators $[T^a,T^b]=if^{abc}T^c$, $g$ is the gauge coupling, $f^{abc}$ denotes the structure constants of $G$, and $m_f$ the fermion masses. In this case, the
 master formula~\eqref{eq:vacuum-trace} gives
\begin{equation}
  \theta^\mu_\mu\big|_\text{YM}
  = \beta_g\,\frac{\partial\mathcal{L}_\text{YM}}{\partial g}
    + \sum_{f=1}^{N_f}\beta_{m_f}\,\frac{\partial\mathcal{L}_\text{YM}}{\partial m_f},
  \quad
  \beta_{m_f} = \mu\,\frac{d m_f}{d\mu} = m_f\,\gamma_{m_f},
\end{equation}
so that~\cite{Collins:1976yq,Nielsen:1977sy}
\begin{equation}
  \theta^\mu_\mu\big|_\text{YM}
  = \frac{\beta_g}{2g}\,F^a_{\mu\nu}F^{\mu\nu}_a
    + \sum_{f=1}^{N_f}m_f\big(1-\gamma_{m_f}\big)\,\bar\psi_f\psi_f
\end{equation}
and
\begin{equation}
\label{trace_anomaly_YM}
  \theta^\mu_\mu\big|_\text{YM}
  = -\frac{g^2}{96\pi^2}\Big[11 C_2(G)-4T(R)\,N_f\Big]\,
    F^a_{\mu\nu}F^{\mu\nu}_a
  +\sum_{f=1}^{N_f}m_f\,\bar\psi_f\psi_f
  + {\cal O}\left(g^4,m_f g^2\right),
\end{equation}
where $C_2(G)$ is the quadratic Casimir operator of the adjoint representation defined by $f^{acd}f^{bcd}=C_2(G)\delta^{ab}$ and $T(R)$ is the trace normalization of the representation $R$, defined by $\text{Tr}_R(T^aT^b)=T(R)\delta^{ab}$ (for $G=\text{SU($N_c$)}$ and $R$ the fundamental representation one has $C_2(G)=N_c$ and $T(R)=1/2$).
For pure Yang--Mills theory ($N_f=0$), asymptotic freedom ($\beta_g<0$) yields a negative gauge‐field coefficient, and the fermion mass term again induces an explicit breaking of scale invariance.

\subsection{Quark-mass contributions and scheme definitions}

Specializing to the case of QCD with $N_f$ quark flavors and strong coupling $g_s$, $\alpha_s=g_s^2/(4\pi)$, the trace anomaly takes the form
\begin{equation}
\label{trace_anomaly_QCD}
 \theta^\mu_\mu\big|_\text{QCD}=\frac{\beta_{g_s}}{2g_s}F_{\mu\nu}^a F^{\mu\nu}_a+\sum_{q=u,d,\ldots}\big(1-\gamma_{m_q}\big)m_q\bar q q
 =-\frac{\alpha_s}{24\pi}\big(33-2N_f\big)F_{\mu\nu}^a F^{\mu\nu}_a+\sum_{q=u,d,\ldots}m_q\bar q q+{\mathcal O}(\alpha_s^2,m_q\alpha_s).
\end{equation}
Based on this expression, we can now proceed with defining more concrete schemes for the mass decomposition as anticipated in Sec.~\ref{intro}. First, for the pion mass, we see that the $\sigma$-term contribution from Eq.~\eqref{M_pi_sigma} encapsulates the matrix element of the quark-mass operator without the anomalous dimension, i.e., assuming that the matrix elements of $q=s,c,b,t$ are negligible, we can write
\begin{equation}
 \mpi=\sigma_\pi+\frac{1}{2\mpi}\Big\langle\pi\Big|\frac{\beta_{g_s}}{2g_s}F_{\mu\nu}^a F^{\mu\nu}_a-\gamma_{m_u}m_u\bar u u-\gamma_{m_d}m_d\bar d d\Big|\pi\Big\rangle,
\end{equation}
which shows that the second half of the pion mass originates from the matrix element of the quantum corrections. Of course, be it via the explicit dependence on $m_q$ or the implicit quark-mass dependence of the states, also this second half has to vanish in the chiral limit, in accordance with Eq.~\eqref{GOR}. One possible scenario for an explicit realization could be an infrared fixed point~\cite{Crewther:2013vea,Zwicky:2023bzk}, at which $\beta_{g_s}=0$ and thus, by consistency with Eq.~\eqref{GOR}, $\gamma_{m_u}=\gamma_{m_d}=-1$~\cite{Zwicky:2023bzk,Zwicky:2023krx}.

For the nucleon, we will develop the analogous decomposition in Sec.~\ref{sec:mass_nucleon}, focusing on the $\sigma$-term contributions
\begin{equation}\label{eq:sigma-terms_def}
 \sigma_{\pi N}=\frac{1}{2\mN}\langle N|m_u\bar u u+m_d\bar d d|N\rangle,\qquad \sigma_q=\frac{1}{2\mN} \langle N|m_q\bar q q|N\rangle\qquad \text{for}\quad q=s,c,b,t.
\end{equation}
Subtracting these contributions, as well as some higher-order quark-mass effects in the case of $q=u,d$, we will provide an estimate of how much of the nucleon mass originates from the different quark flavors and the gluon fields, respectively, see Refs.~\cite{Bali:2016lvx,Liu:2021gco,Liu:2023cse} for similar decompositions. A further breakdown often employed in the literature amounts to writing~\cite{Bali:2016lvx,Liu:2021gco,Liu:2023cse,Tarrach:1981bi,Ji:1994av}
\begin{equation}
\label{decomp_kin_a}
M_H = \langle H| {\mathcal H}_{m_q}|H\rangle+\langle H| {\mathcal H}_\text{kin}|H\rangle+\langle H| {\mathcal H}_{a}|H\rangle,
\end{equation}
where ${\mathcal H}_{m_q}$ gives rise to the quark-mass contribution determined by the sum of the $\sigma$-terms, ${\mathcal H}_\text{kin}$ describes the quark and gluon kinetic energies, and ${\mathcal H}_{a}$ is the remainder of the trace anomaly. Since at zero momentum $\langle H| {\mathcal H}_\text{kin}|H\rangle=3\langle H| {\mathcal H}_{a}|H\rangle$~\cite{Bali:2016lvx} and thus $\langle H| {\mathcal H}_{a}|H\rangle=(M_H-\langle H| {\mathcal H}_{m_q}|H\rangle)/4$, this decomposition can be easily constructed from the various $\sigma$-terms as discussed in Sec.~\ref{sec:mass_nucleon}, and hence we will not dwell on Eq.~\eqref{decomp_kin_a} further. Finally, we mention calculations in QCD sum rules as another complementary perspective on hadron masses~\cite{Ioffe:1981kw}.

\section[Nucleon $\sigma$-terms]{Nucleon $\boldsymbol{\sigma}$-terms}
\label{sec:mass_nucleon}

The nucleon $\sigma$-terms encapsulate the contributions to the nucleon mass that originate from explicit breaking of scale invariance by the quark masses. 
The isospin-symmetric light-quark combination
$\sigma_{\pi N}$, as defined in Eq.~\eqref{eq:sigma-terms_def}, governs most of the up- and down-quark content of $\mN$ and, by virtue of a low‐energy theorem, is directly related to the isospin‐even $\pi N$ scattering amplitude at the unphysical Cheng--Dashen point.  The strange $\sigma$-term $\sigma_{s}$ probes the hidden-strangeness contribution to the nucleon's mass, as may arise due to virtual $\bar s s$ pairs even though the nucleon does not possess strange valence quarks. 
In the same way, heavy‐quark $\sigma$-terms quantify how virtual charm-, bottom-, and top-quark pairs influence $\mN$, in this case, mostly via their coupling to gluons and the large gluonic nucleon matrix element.

In this section, we summarize the present knowledge of all these matrix elements, starting with the phenomenological determination of $\sigma_{\pi N}$ based on the interplay of the Cheng--Dashen low-energy theorem and a precision calculation of $\pi N$ scattering using Roy--Steiner equations, see Sec.~\ref{sec:sigma_piN_pheno}. Drawing conclusions on the decomposition of $\mN$ also requires a brief discussion of isospin conventions and the extrapolation to the chiral limit, as provided in Sec.~\ref{sec:isospin}.
Alternatively, in lattice QCD, one may compute $\sigma_{\pi N}$ directly via three‑point correlators of the scalar density in Eq.~\eqref{eq:sigma-terms_def} or indirectly by applying the Feynman--Hellmann theorem to the quark‑mass dependence of $\mN$, as spelled out in Eq.~\eqref{eq:FH_def}. In contrast to $\sigma_{\pi N}$, a robust phenomenological determination of $\sigma_s$ with controlled uncertainties is not possible; in Sec.~\ref{sec:SU3} we lay out the reasoning why that is the case. Accordingly, $\sigma_{\pi N}$ becomes an important test case, allowing for a benchmark of lattice-QCD and phenomenological evaluations, the status of which is reviewed in Sec.~\ref{sec:lattice} together with current results for $\sigma_s$ and $\sigma_c$. Finally, the determination of $\sigma_{q}$, $q=c,b,t$, with perturbative methods is discussed in Sec.~\ref{sec:pQCD}.

\subsection[Phenomenological determinations of $\sigma_{\pi N}$]{Phenomenological determinations of $\boldsymbol{\sigma_{\pi N}}$}
\label{sec:sigma_piN_pheno}

The phenomenological determination of $\sigma_{\pi N}$ relies on the Cheng--Dashen low‐energy theorem~\cite{Cheng:1970mx,Brown:1971pn}, which relates the scalar form factor of the nucleon,
\begin{equation}
\sigma(t)
=\frac{1}{2\mN}\langle N(p')|\hat m(\bar u u+\bar d d)|N(p)\rangle,\qquad t=(p'-p)^2,
\end{equation}
evaluated at $t=2\mpi^2$, to the Born-term-subtracted isoscalar $\pi N$ scattering amplitude  $\bar D^+(\nu,t)$, evaluated at the Cheng--Dashen point $(\nu=0,t=2\mpi^2)$ ($\nu$ is the crossing-symmetry-odd combination of Mandelstam variables for $\pi N$ scattering, $t$ the invariant mass of the $\pi\pi\to\bar NN$ reaction). This point lies in the unphysical region, i.e., outside the region that can be directly accessed by $\pi N$ scattering experiments, but the corresponding value can be obtained by means of analytic continuation. Accordingly, one has the relation  
\begin{equation}
\label{LET}
\bar D^+(0,2\mpi^2)=\sigma(2\mpi^2)+\Delta_R,
\end{equation}
where $\Delta_R$ encodes higher-order corrections in ChPT. In fact, the low-energy theorem is particularly stable, as the expansion proceeds in SU(2), with corrections suppressed by powers of $\mpi$, and chiral logarithms or other non-analytic effects that might slow down the convergence are absent at full one-loop order~\cite{Bernard:1996nu,Becher:2001hv}, leading to the estimate  
$|\Delta_R|\lesssim 2\MeV$. In practice, the Cheng--Dashen theorem is recast into the form
\begin{equation}\label{LET_v2}
\sigma_{\pi N}=\sigma(0)=\Sigma_d+\Delta_D-\Delta_\sigma-\Delta_R,
\end{equation}
where $\Sigma_d = F_\pi^2(d_{00}^+ + 2\mpi^2\,d_{01}^+)$ is the leading term in the subthreshold expansion of the $\pi N$ amplitude, $\Delta_D$    
accounts for higher-order terms in the isoscalar amplitude beyond $d_{00}^+$ and $d_{01}^+$, and $\Delta_\sigma = \sigma(2\mpi^2)-\sigma(0)$ measures the curvature of the scalar nucleon form factor between $t=0$ and $t=2\mpi^2$.
Although $\Delta_D$ and $\Delta_\sigma$ are individually large---driven by strong isospin-$0$ $\pi\pi$ $S$‐wave rescattering~\cite{Donoghue:1990xh}---their difference largely cancels~\cite{Gasser:1990ap}, and the most precise determination to date~\cite{Hoferichter:2012wf}, also including $\bar K K$ intermediate states, finds $\Delta_D-\Delta_\sigma=-1.8(2)\MeV$.
In this way, the remaining information on the scattering amplitude is encoded in the subthreshold parameters $d_{00}^+$ and $d_{01}^+$, whose determination requires the aforementioned analytic continuation of the $\pi N$ isoscalar amplitude to the Cheng--Dashen point. 

To control this analytic continuation, one relies on dispersion relations, with input determined in the physical region. Such evaluations have been performed, e.g., in  Refs.~\cite{Koch:1980ay,Hohler:1984ux,Gasser:1988jt,Gasser:1990ce,Pavan:2001wz,Arndt:2006bf,Workman:2012hx}, based on partial-wave analyses of $\pi N$ scattering to constrain the input in the dispersion relations (direct analyses in ChPT tend to be consistent with the partial-wave analysis used to determine the low-energy constants~\cite{Fettes:2000xg,Alarcon:2011zs}). However, given the fact that relatively few measurements are available in the critical low-energy region, additional constraints are required for a precision determination of $\sigma_{\pi N}$. This program, started in Ref.~\cite{Ditsche:2012fv}, involved setting up and solving partial-wave dispersion relations for $\pi N$ scattering~\cite{Ditsche:2012fv} in the form of Roy--Steiner equations~\cite{Baacke:1970mi,Steiner:1970mh,Steiner:1971ms,Hite:1973pm}, as well as the evaluation of isospin-breaking corrections to the low-energy theorem~\cite{Hoferichter:2015dsa}. Apart from the determination of $\sigma_{\pi N}$~\cite{Hoferichter:2015dsa,Hoferichter:2016ocj,RuizdeElvira:2017stg,Hoferichter:2023ptl}, this program~\cite{Hoferichter:2015hva} has led to 
 a number of other applications, including low-energy constants in ChPT~\cite{Hoferichter:2015tha,Siemens:2016jwj}, nucleon form factors~\cite{Hoferichter:2016duk,Hoferichter:2018zwu,Crivellin:2023ter,Cao:2024zlf}, and resonance properties~\cite{Hoferichter:2023mgy}.

A second important development in the phenomenological determination of $\sigma_{\pi N}$ concerns experimental information on the low-energy $\pi N$ scattering amplitude. Given the scarce data situation from scattering experiments, measurements in pionic atoms~\cite{Strauch:2010vu,Hennebach:2014lsa,Hirtl:2021zqf} (bound systems of a $\pi^-$ and a proton or a deuteron) have proved particularly valuable, providing independent constraints on the $\pi N$ scattering lengths. Once the required isospin-breaking~\cite{Gasser:2002am,Gasser:2007zt,Hoferichter:2009ez,Hoferichter:2009gn} and few-body~\cite{Weinberg:1992yk,Beane:2002wk,Baru:2004kw,Lensky:2006wd,Baru:2007wf,Liebig:2010ki,Baru:2012iv} corrections are applied, a combined analysis of the available measurements~\cite{Baru:2010xn,Baru:2011bw} yields the necessary input to determine the free parameters in the solution of the Roy--Steiner equations, leading to the final result~\cite{Hoferichter:2023ptl}
\begin{equation} 
\label{pheno_final}
    \sigma_{\pi N}=59.0(3.5)\MeV.
\end{equation}
A consistent, albeit less precise result, $\sigma_{\pi N}=58(5)\MeV$, was derived in Ref.~\cite{RuizdeElvira:2017stg} using low-energy cross-section data (including relatively recent data on elastic reactions~\cite{Brack:1989sj,Joram:1995gr,Denz:2005jq} and the charge-exchange process~\cite{Frlez:1997qu,Isenhower:1999aj,Jia:2008rt,Mekterovic:2009kw}) to determine the free parameters in the Roy--Steiner solution instead of pionic atoms, confirming Eq.~\eqref{pheno_final} at the level of $5\MeV$. Equation~\eqref{pheno_final} is also consistent with an extraction of $\sigma_{\pi N}$ from in-medium modifications of the isovector scattering length in pionic atoms with large $Z$~\cite{Weise:2000xp,Weise:2001sg,Friedman:2019zhc}, but the nuclear uncertainties in such determinations are substantial.

\subsection{Isospin conventions and chiral limit}
\label{sec:isospin}

As discussed in detail in Ref.~\cite{Hoferichter:2023ptl}, the definition of $\sigma_{\pi N}$ depends on the conventions for the isospin limit. Here, we choose the mass of the charged pion, following the conventions established in Ref.~\cite{Meissner:1997ii}, motivated by the fact that data input for the phenomenological analysis is primarily available for processes involving charged particles. In contrast, lattice-QCD calculations favor a definition in terms of the mass of the neutral pion, see Sec.~\ref{sec:lattice}, so that a correction $\Delta\sigma_{\pi N}=\sigma_{\pi N}-\bar\sigma_{\pi N}=3.1(5)\MeV$~\cite{Hoferichter:2023ptl} needs to be applied to compare $\bar\sigma_{\pi N}$ from lattice QCD to $\sigma_{\pi N}$. Isospin-breaking corrections can be added to Eq.~\eqref{pheno_final} to obtain the scalar couplings in a particular channel, for $q=u,d$ in proton or neutron~\cite{Crivellin:2013ipa,Hoferichter:2023ptl}, where input for the additionally required matrix elements needs to be provided from lattice QCD~\cite{Giusti:2017dmp,MILC:2018ddw,FlavourLatticeAveragingGroupFLAG:2021npn,Borsanyi:2014jba,Brantley:2016our,CSSM:2019jmq} and/or phenomenology~\cite{Gasser:1974wd,Gasser:2015dwa,Gasser:2020mzy,Gasser:2020hzn}. However, here we restrict the presentation to the isospin limit as shown in Eq.~\eqref{pheno_final}.  

Returning to the decomposition of $\mN$, $\sigma_{\pi N}$ quantifies the direct contribution of the $q=u,d$ scalar current, but does not yet include the effects of higher orders in the chiral expansion, i.e., does not account for the dependence of the other terms in the nucleon matrix element of Eq.~\eqref{trace_anomaly_QCD} on $\hat m$. To estimate the corresponding corrections, we consider the chiral expansion of the nucleon mass up to fourth order 
\begin{equation}
\label{mN_chiral}
\mN=\tilde m^{(2)}_N-4c_1\mpi^2-\frac{3g_A^2\mpi^3}{32\pi\Fpi^2}-\frac{3}{32\pi^2\Fpi^2\mN}\Big[g_A^2+\mN\big(-8c_1+c_2+4c_3\big)\Big]\mpi^4\log\frac{\mpi}{\mN}+\bigg[e_1-\frac{3\big(2g_A^2-c_2\mN\big)}{128\pi^2\Fpi^2\mN}\bigg]\mpi^4+{\mathcal O}\big(\mpi^5\big),
\end{equation}
where $g_A$ is the axial charge of the nucleon, $c_{1,2,3}$ are low-energy constants that can also be determined from $\pi N$ scattering~\cite{Hoferichter:2015tha}, and $e_1$ denotes an additional low-energy constant that follows when imposing $\sigma_{\pi N}$ from Eq.~\eqref{pheno_final}. Moreover, various isospin-breaking effects have been absorbed into $\tilde m^{(2)}_N$~\cite{Hoferichter:2015hva}, but otherwise this parameter quantifies how much of the nucleon mass remains if $\hat m\to 0$ (in general, we denote by $\tilde m_N^{(N_f)}$ the nucleon mass after $N_f$ quark flavors have been removed). The result, $\tilde m^{(2)}_N\simeq 869.5\MeV$, shows that the entire contribution from $q=u,d$ amounts to about $69\MeV$, thus $\simeq 10\MeV$ more than from $\sigma_{\pi N}$ alone. Most of this difference can be understood from the third-order term in Eq.~\eqref{mN_chiral}, since
\begin{equation}
\label{third_order_SU2}
 m_N-\sigma_{\pi N} -\tilde m^{(2)}_N = \frac{3g_A^2\mpi^3}{64\pi\Fpi^2}+{\mathcal O}\big(\mpi^4\big)\simeq 8\MeV  +{\mathcal O}\big(\mpi^4\big),   
\end{equation}
which thus almost saturates the total amount of higher-order corrections (for the pion, the analogous relation would be $\mpi-\sigma_{\pi}=\mpi/2+{\mathcal O}(\mpi^3)$). According to this discussion, we will use 
\begin{equation}
\label{mN_ud}
    \mN^{ud}\simeq 69\MeV
\end{equation}
as an estimate for the contribution to the nucleon mass that originates from the light quark masses $m_u$ and $m_d$.

\subsection[SU(3) relations for $\sigma_s$]{SU(3) relations for $\boldsymbol{\sigma_s}$}
\label{sec:SU3}

While the low-energy theorem for $\sigma_{\pi N}$ is protected by SU(2) flavor symmetry, rendering its extraction from $\pi N$ scattering rather robust, the strange $\sigma$-term $\sigma_s$ can only be determined phenomenologically when relying on SU(3) symmetry~\cite{Gasser:1980sb,Gasser:1982ap,Gasser:1990ce}. To be concrete, at leading order in the SU(3) expansion one has 
\begin{equation}
\sigma_0\equiv\frac{\hat m}{2\mN}\langle N|\bar u u+\bar d d-2\bar s s|N\rangle = \frac{\hat m}{m_s-\hat m}  
\left(m_\Xi + m_\Sigma - 2\mN\right)\simeq 26\MeV,
\end{equation}
a value that is increased to $\sigma_0=35(5)\MeV$ when considering higher-order corrections~\cite{Gasser:1982ap}. However, even this increased value would suggest an enormous strange $\sigma$-term 
\begin{equation}
 \sigma_s=\frac{m_s}{2\hat m}\big(\sigma_{\pi N}-\sigma_0\big)\simeq 330\MeV,    
\end{equation}
where we used $m_s/\hat m\simeq 27.3$~\cite{MILC:2009ltw,BMW:2010ucx,BMW:2010skj,EuropeanTwistedMass:2014osg,FermilabLattice:2014tsy,RBC:2014ntl,Bazavov:2017lyh,Bruno:2019xed,ExtendedTwistedMass:2021gbo,FlavourLatticeAveragingGroupFLAG:2024oxs}, completely at odds with the findings in lattice QCD, see Sec.~\ref{sec:lattice}. The solution to this puzzle is that even in the SU(2) sector the convergence of heavy-baryon ChPT can be challenging, casting doubt on the viability of SU(3) expansions in the baryon sector. That is, while corrections to the Cheng--Dashen theorem are particularly well behaved, there are cases in which enhancements due to $\Delta$ degrees of freedom, by numerical factors of $\pi$, or large logarithmic corrections render the convergence properties of SU(2) observables challenging, e.g., (sub)threshold parameters in $\pi N$ scattering~\cite{Siemens:2016jwj} or the axial charge of the nucleon~\cite{Hall:2025ytt}. In fact, one such enhancement by a factor of $\pi$ already appeared in Eq.~\eqref{third_order_SU2}, tracing back to the particular loop function, but in this case $\mpi$ is small enough that the expansion still converges well, as demonstrated by the small ${\mathcal O}(\mpi^4)$ term. It is instructive to consider the same effect in the SU(3) expansion after $\sigma_s$ has been subtracted, since in this case $\mN$ again receives a non-analytic loop correction that is predicted in terms of the known SU(3) parameters $D\simeq 0.8$, $F\simeq 0.46$. The result~\cite{Mai:2009ce,Frink:2004ic,Borasoy:1996bx}
\begin{equation}
\mN-\mN^{ud}-\sigma_s-\tilde m^{(3)}=\frac{M_K^3}{96\pi\Fpi^2}\bigg[5D^2-6DF+9F^2+\frac{4}{3\sqrt{3}}\big(D-3F\big)^2\bigg]+{\mathcal O}\big(M_K^4\big)\simeq  150\MeV +{\mathcal O}\big(M_K^4\big),   
\end{equation}
is again huge, illustrating how the sizable kaon mass in combination with $\pi$ enhancements and large coefficients dispels any hope that the SU(3) expansion may converge.  For that reason, we will simply use $\sigma_s$ as determined in lattice QCD, to which we will turn next.   

\subsection{Lattice QCD}
\label{sec:lattice}

In lattice QCD, nucleon $\sigma$-term matrix elements can be determined using two different strategies, either by calculating the quark-mass derivative of the nucleon mass and applying the Feynman--Hellmann theorem~\eqref{eq:FH_def} or by direct calculation of the three-point function of a nucleon source, sink, and insertion of a scalar current. In both methods, the uncertainties of the resulting $\sigma$-terms comprise statistical errors, which become most challenging for light quarks and so-called disconnected diagrams, and systematic errors, reflecting, e.g., the uncertainties related to the extrapolation to the continuum limit, physical quark masses, and infinite volume. In addition, nucleon correlation functions can be affected by excited-state contamination, essentially, contributions from a tower of excited states with the same quantum numbers as the nucleon. A comprehensive review of the status of the resulting $\sigma$-terms is provided in Ref.~\cite{FlavourLatticeAveragingGroupFLAG:2024oxs}, including averages of those calculations that pass quality criteria related to the aforementioned systematic uncertainties. These averages are collected in Table~\ref{tab:FLAG}, with separate entries for $N_f=2+1$ and $N_f=2+1+1$ quark flavors (corresponding to $q=u, d, s$ or $q=u, d, s, c$ active quark flavors, and isospin-limit for $q=u,d$). 

\begin{table}[t]
	\TBL{\caption{Nucleon $\sigma$-terms from lattice QCD, as compiled in Ref.~\cite{FlavourLatticeAveragingGroupFLAG:2024oxs}.}}{%
		\begin{tabular*}{\columnwidth}{@{\extracolsep\fill}lrrrr}
			\toprule
			 & $\bar \sigma_{\pi N} \, [\text{MeV}]$ & References & $\sigma_s \, [\text{MeV}]$& References\\
			\colrule
			$N_f=2+1$ & $42.2(2.4)$ & Refs.~\cite{BMW:2011sbi,Durr:2015dna,Yang:2015uis,RQCD:2022xux,Agadjanov:2023efe} & $44.9(6.4)$  & Refs.~\cite{BMW:2011sbi,Durr:2015dna,Yang:2015uis,RQCD:2022xux,Agadjanov:2023efe,Freeman:2012ry,Junnarkar:2013ac}\\
            $N_f=2+1+1$ & $60.9(6.5)$& Refs.~\cite{Alexandrou:2014sha,Gupta:2021ahb} & $41.0(8.8)$ & Ref.~\cite{Freeman:2012ry}\\
			\botrule
	\end{tabular*}}{\label{tab:FLAG}}
\end{table}

In the case of $\bar\sigma_{\pi N}$, the $N_f=2+1$ results display a $3.2\sigma$ tension with Eq.~\eqref{pheno_final}, whose origin is not fully understood at present. As argued in Refs.~\cite{Gupta:2021ahb,Gupta:2022aba}, a possible resolution at least for the direct three-point-function method could concern the role of excited-state contamination~\cite{Bar:2016uoj}. In fact, the preferred analysis in Ref.~\cite{Gupta:2021ahb} imposes priors on the excited-state spectrum motivated by ChPT, leading to a significant enhancement of $\sigma_{\pi N}$, as reflected by the $N_f=2+1+1$ value quoted in Table~\ref{tab:FLAG}. The calculation in Ref.~\cite{Agadjanov:2023efe} also suggests that the treatment of excited states may be important, but in neither case the sensitivity of the lattice data suffices yet to extract the excited-state spectrum at the required level of precision. Moreover, reconciling a value of $\bar\sigma_{\pi N}\simeq 55\MeV$ with Feynman--Hellmann calculations remains an open question (see, e.g., Refs.~\cite{Chen:2012nx,Ren:2016aeo,Ren:2017fbv,Lutz:2018cqo,Lutz:2023xpi} for recent analyses in ChPT). In the current situation, we will adopt the estimate quoted in Eq.~\eqref{mN_ud}, keeping in mind that the precise value of the light-quark contribution to $\mN$ is not yet fully settled.

For $\sigma_s$, we already argued in the previous section that there is no way to extract its value from phenomenology in a reliable manner, with SU(3)-based determinations predicting values much bigger than the lattice-QCD results  given in Table~\ref{tab:FLAG}. Since the latter agree reasonably well among the different calculations (as well as between the $N_f=2+1$ and $N_f=2+1+1$ averages), we will use the corresponding average
\begin{equation}
\label{mN_s}
    m_N^s\simeq 43\MeV
\end{equation}
as an estimate of the contribution to the nucleon mass generated by $m_s$. In principle, there could again be higher-order effects as in the case of $m_N^{ud}$, but the expectation would be that such corrections will become less relevant for the non-valence quarks.  

For $\sigma_c$, no averages are provided in Ref.~\cite{FlavourLatticeAveragingGroupFLAG:2024oxs}, as  too few calculations exist that pass the quality criteria, leaving no meaningful averages for the different $N_f$ values. However, all available calculations are consistent with the range $\sigma_c=80(20)\MeV$~\cite{Bali:2016lvx,Freeman:2012ry,XQCD:2013odc,Abdel-Rehim:2016won,Alexandrou:2019brg,Borsanyi:2020bpd}, which we can now compare to perturbative arguments for the heavy quarks $Q=c,b,t$.

\subsection{Heavy quarks}
\label{sec:pQCD}

For the heavy quarks, $Q=c,b,t$, the matrix elements $\langle N|m_Q\bar Q Q|N\rangle$ can be studied with perturbative techniques~\cite{Shifman:1978zn}. The basic idea is that ultimately these scalar couplings are obtained by integrating out a heavy-quark loop coupled to gluons, i.e., from Eq.~\eqref{trace_anomaly_QCD} it follows that when integrating out a heavy quark from $N_f+1$ to $N_f$, the matrix element needs to fulfill
\begin{equation}
\label{heavy_quark}
\langle N|m_Q\bar Q Q|N\rangle =-\frac{\alpha_s}{12\pi} \langle N|F_{\mu\nu}^aF^{\mu\nu}_a|N\rangle =\frac{2}{27}\Big(\mN-\sigma_{\pi N}-\sigma_s\Big)\simeq 62\MeV,
\end{equation}
to ensure that Eq.~\eqref{trace_anomaly_QCD} continues to hold, where the second step is obtained by considering  
 Eq.~\eqref{trace_anomaly_QCD} with $N_f=3$ and both identities are valid at leading order in $\alpha_s$. Corrections to Eq.~\eqref{heavy_quark} have been worked out up to ${\mathcal O}(\alpha_s^3)$ \cite{Chetyrkin:1997un,Hill:2014yxa}, but given the substantial uncertainties in the $q=u,d,s$ $\sigma$-terms we restrict ourselves the the leading $\alpha_s$ corrections~\cite{Hoferichter:2017olk}, which gives 
 \begin{equation}
 \label{mN_Q}
\sigma_c\simeq 68\MeV,\qquad \sigma_b\simeq 65\MeV,\qquad \sigma_t\simeq 63\MeV.
 \end{equation}
 For $\sigma_c$, the perturbative result therefore agrees with current lattice-QCD estimates, but it would be important to benchmark $\sigma_c$ at a higher level of precision.  

\section{Summary of the nucleon mass decomposition}
\label{sec:summary}

Taking together the numerical estimates for the various quark-mass  contributions from Eqs.~\eqref{mN_ud}, \eqref{mN_s}, and~\eqref{mN_Q}, we obtain for the decomposition of the nucleon mass
\begin{equation}
\label{mN_numerics}
 \mN=\mN^{ud}+\mN^s+\mN^c+\mN^b+\mN^t+\mN^{FF}
 \simeq \Big(69+43+68+65+63+630\Big)\MeV,
\end{equation}
leading to the picture already anticipated in Fig.~\ref{fig:mN_decomposition}. In particular, the implication is that $\mN^{FF}\simeq 630\MeV$ originating from the gluon field strength would remain in the absence of quark-mass contributions, so that two thirds of the nucleon mass are of gluonic origin, about $21\%$ can be identified with the heavy quarks, $5\%$ with the strange quark, and only $7\%$ with the valence quarks of the nucleon. We emphasize that the precise values rely on a number of assumptions, e.g., we used the phenomenological value for $\sigma_{\pi N}$ from Eq.~\eqref{pheno_final} (including higher-order quark-mass corrections), neglected such corrections for the strange quark, and used perturbative input for $\sigma_c$.  However, while the precise numerical breakdown in Eq.~\eqref{mN_numerics} may change with future improved evaluations of the corresponding matrix elements, the hierarchy among the different contributions is robust, as illustrated in Fig.~\ref{fig:mN_decomposition}.

\section{Conclusions}
\label{sec:conclusions}

In this chapter, we first provided an introduction to the trace anomaly of the energy-momentum tensor, $\theta^\mu_\mu$, in particular, its relation to the origin of hadron masses. While the structure of $\theta^\mu_\mu$ is rather general, the application to hadron masses is based on the special case for QCD as given in Eq.~\eqref{trace_anomaly_QCD}, decomposing into anomalous contributions that only appear in the quantum theory and an explicit violation of scale invariance due to the quark masses. One can then ask the question how much of the mass of a given hadron traces back to the different terms in $\theta^\mu_\mu$. 

Even for the example of the pion, whose mass vanishes in the chiral limit $m_{u,d}\to 0$, this is a non-trivial question, as the direct, $\sigma$-term contribution from the $m_q\bar q q$ operator only generates $50\%$ of its mass, while the remainder is related to quantum corrections. The case of the nucleon is even more interesting, as its mass stays finite in the chiral limit and also quarks other than $q=u,d$ play a role, either because the direct $\bar q q$ matrix is non-negligible (as for $q=s$) or because it is related to the matrix element of gluon field strength operator $F_{\mu\nu}^aF^{\mu\nu}_a$ (as for $q=c,b,t$). Accordingly, it is a priori not clear how much the individual terms contribute. In this chapter, we provided our current best estimates, based on (i) phenomenological determinations of the $\sigma$-term for $q=u,d$, $\sigma_{\pi N}$, via the Cheng--Dashen low-energy theorem as well as chiral corrections to the nucleon mass, (ii) lattice QCD for $q=u,d,s,c$, and (iii) perturbation theory for $q=c,b,t$. The resulting picture is summarized in Fig.~\ref{fig:mN_decomposition}, with about two thirds from gluonic origin, $21\%$ from the heavy quarks, $5\%$ from the strange quark, and only $7\%$ from the valence quarks. However, while the general hierarchy is robust, the individual numbers may change with future improved evaluations of the corresponding matrix elements. In this regard, the main open questions are: 
\begin{itemize}
	\item What is the origin of the tension between $\sigma_{\pi N}$ determinations from lattice QCD and phenomenology? Can excited-state contamination explain the difference?
    \item Are there relevant higher-order corrections in the strangeness chiral limit $m_s\to 0$ not captured by $\sigma_s$?
    \item How reliable is perturbation theory for $\sigma_c$?
\end{itemize}

\begin{ack}[Acknowledgments]%
We thank Bastian Kubis and Ulf-G.~Mei{\ss}ner for decade-long collaboration on the subject of this chapter.
 Financial support by the SNSF (Project No.\ TMCG-2\_213690), the Ramon y Cajal program (RYC2019-027605-I) of
the Spanish MICIU, and the Spanish Ministerio de Ciencia e Innovacion (Project PID2022-136510NB-C31) is gratefully acknowledged.
\end{ack}


\bibliographystyle{elsarticle-num}
\bibliography{reference.bib}

\end{document}